\documentstyle[aps,subary,psfig]{revtex}

\newcommand{\iy}{\infty}
\newcommand{\be}{\begin{equation}}
\newcommand{\ee}{\end{equation}}
\newcommand{\bex}{\begin{eqnarray}}
\newcommand{\eex}{\end{eqnarray}}

\begin{document}

\title{Noncausal Superluminal Nonlocal Signalling}
\author{R. Srikanth\thanks{e-mail: srik@iiap.ernet.in}\\ 
Indian Institute of Astrophysics, India.}
\maketitle
\date{}


\begin{abstract}
We propose a thought experiment 
for classical superluminal signal transmission based on the quantum
nonlocal influence
of photons on their momentum entangled EPR twins. The signal sender
measures either position or momentum of particles in a pure ensemble 
of the entangled pairs, leaving their twins as 
localized particles or plane waves. The signal receiver distinguishes these
outcomes interferometrically using a double slit interferometer modified by
a system of optical filters. Since the collapse of the wavefunction is
postulated to be instantaneous, this signal can be transmitted superluminally.
We show that the method circumvents the no-signalling theorem because  
the receiver is able to modify the disentangled wavefunction before his
measurement. We propose a plan for the possible practical realization of a
superluminal quantum telegraph based on the thought experiment.
\end{abstract}

\section{Introduction}

One of the counterintuitive features of quantum mechanics is nonlocality, 
whereby operations on a system can instantaneously 
affect measurements on a distant system with which it
had interacted in the past but is now no longer in dynamical interaction with. 
It was first considered by Einstein, Podolsky and Rosen (EPR), 
who showed that quantum mechanics is either 
incomplete or nonlocal \cite{einstein}. 
They rejected the nonlocal possibility because it implied a
superluminal influence in apparent contradiction with Special
Relativity. However, today quantum nonlocality 
is widely accepted as a fact thanks to
experiments since the mid-1980's performed both on spin entangled systems
\cite{astiw}, based on the Bohm version \cite{bohm} of the EPR 
thought-experiment, which are shown to violate Bell's inequality
\cite{bella} (apart from technical caveats concerning detection loopholes
\cite{thompson}), and also on systems entangled in continuous variables
(Refs. \cite{ghosh,kwi93,strekalov,zei2000} and references therein), 
where nonlocality is manifested directly in multi-particle interferences. 

It is generally
accepted that although quantum nonlocality enforces distant superluminal
correlations at the level of individual events, the inherent randomness 
of the correlated processes prevents the transmission of
controllable signals \cite{eberhard2}. One sees this by noting that 
the expectation value of an observable of one of a pair of entangled
systems remains unaffected by the interaction
of a measuring apparatus with the other
\cite{bohi,ghirardi,shi93}. Bussey proved that the probability 
for outcomes of a measurement on an entangled particle is not affected by a 
measurement on the other entangled particle \cite{bussey}. Jordan 
argued that the particles of an entangled pair belongs
to different subsystems and hence operations on them should commute
\cite{jordan}. On this
basis, he concluded that expectation value for the given observable at one
of the particles remains unaffected by operations at the other. In the language
of Bayesian probability, Garrett pointed out that the 
{\it a priori} and {\it a posteriori} probabilities of results of 
measurments on a particle are the same given a measurement on its 
entangled counterpart \cite{garrett}.
Using the density matrix formulation to resolve the EPR `paradox', Cantrell 
and Scully proved that the reduced density matrix remains the
same for both particles in an EPR pair before and after measurement 
on one of them \cite{cantrell}. On this basis , it is conventionally accepted 
that that nonlocality and Relativistic causality can ``coexist peacefully"
\cite{shi89}.

However Peacock and Hepburn have argued that most existing 
no-signalling proofs are to an extent tautological since they begin by 
assuming an
explicitly local interaction Hamiltonian between the measuring instrument
and the entangled system \cite{peacock}. Here we prove
by means of a thought experiment that the superluminal nonlocal influence is
in fact noncausal. 
The assumption of instantaneous collapse of the wavefunction throughout
configuration space is thus shown to be not of purely epistemological
significance as held in the orthodox Copenhagen interpretation but as 
leading to controllable superluminal physical effects.

\section{A Thought Experiment}\label{thought}

For the purpose of superluminal communcation, the advantage of entanglement
in a spatial variable like momentum over spin 
is that one can modify the wavefunction by using mirrors,
lenses, filters and double (or multiple) slits before making a
measurement. Such a modification is crucial to avoid the consequences 
of the no-signalling theorem, as we later see. Mirrors and lenses can be
used to rotate momentum states, while slits can be used for getting the
wavefunction to self-interfere-- a feature central to the argument below.

\subsection{The experimental set-up}

\begin{figure}
\centerline{\psfig{file=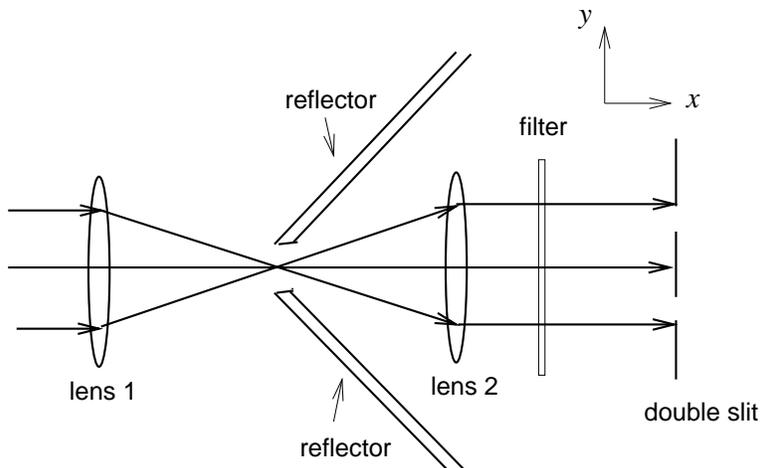}}
\caption{Bob's equipment: the tandem lens system acts as a directional
filter that permits only horizontal rays to fall on the double slit. The 
spectral filter restricts the radiation from lens 2 to a narrow bandwidth. The
entire system is placed within an enclosure with a reflecting exterior,
so that the only radiation entering the experiment is via lens 1.}
\label{bm}
\end{figure}

The proposed thought experiment is a modification of the original
Einstein-Podolsky-Rosen experiment \cite{einstein}.
It involves source S of light consisting of
entangled photon-pairs. Two spatially seperated observers,
Alice and Bob analyze the light. Alice is equipped with a device, labelled
K, to measure
momentum or position of the photons in $x$ and $y$ direction. Bob is
equipped with a Young's double-slit interferometer, labelled L,
with the slits seperated along the $y$-axis. A directional filter assembly
is mounted just before the double slit diaphragm, as shown in Figure \ref{bm},
to ensure that only rays (almost) orthogonal to its surface will fall on it.
It consists of two convex lenses in tandem such that their foci coincide.
Entry of light towards lens 2 is restricted to
a small hole in the focal plane surrounding the focus. 
All other light passing from the
lens 1 is reflected backwards by the reflector. A spectral filter
behind lens 2 is used to restrict the radiation to a narrow bandwidth
about wavelength $\lambda$. The double slit is shielded
from stray radiation by an enclosure with a
reflecting exterior. Thus, the only radiation entering the experiment is
via lens 1. Together
the filters and the enclosure pass only a unidirectional almost
monochromatic light onto the double slit. 
The use of reflectors rather than absorbers is to
avoid collapsing the wavefunction.
 
All measurements and detectors are considered
to be ideal. Part of the light (denoted A) from S is analyzed by Alice at
time $t_1$ while its EPR counterpart (denoted B) is analyzed by Bob at 
$t_2 \equiv t_1 + \delta t$, where $\delta t$ includes the time taken by
Bob to complete his observation. 
By prior agreement, at $t_1$ Alice will measure either the 
transverse position coordinate $y$ and longitudinal momentum componant
$p_x$ of the photons in ensemble A, or the momentum componants $p_x, p_y$,
while Bob will observe the resulting interference pattern on the screen of L.
Just before her measurement, only Alice knows what she will measure.
We shall prove that quantum nonlocality implies that at $t_2$ Bob can
know what measurement Alice has made.

\subsection{Nonlocal communication}

At time $t_1$ let the pure state of the entangled EPR 
pairs be described by the wavefunction:
\begin{equation}
\label{wavefunc1}
\Psi ({\bf x}_a, {\bf x}_b) = \int\int_{-\iy}^{\iy}
 e^{(2\pi i/\hbar )({\bf x}_a - {\bf x}_b + {\bf x_0})\cdot{\bf p}}dp_xdp_y,
\end{equation}
where ${\bf x} \equiv (x, y)$, the
subscripts $a$ and $b$ refer to the particles in ensembles A and B,
respectively, ${\bf x}_0$ is the Alice-Bob spatial seperation, 
and ${\bf p} \equiv (p_x, p_y)$ is momentum.
At time $t_1$, Alice has the choice of making either a transverse position 
or momentum measurement on the particles in A. 
The corresponding operators are $y$ and 
$(\hbar /2\pi i)\nabla$.
If Alice measures the momentum, then a photon in A is
collapsed to a momentum eigenstate:
\begin{equation}
\label{eigenmom1}
\pi_a({\bf x}_a) = e^{(2\pi i/\hbar){\bf x_a}\cdot{\bf p}}
\ee
with the eigenvalue of ${\bf p} = (p_x, p_y)$.
Simultaneously, the photon B is left in the momentum eigenstate:
\begin{equation}
\label{eigenmom2}
\pi_b({\bf x}_b) = e^{(2\pi i/\hbar)({\bf x}_0 - {\bf x}_b)\cdot{\bf p}},
\ee
with eigenvalue $-{\bf p}$.
Eq. (\ref{wavefunc1}) can now be written as:-
\be
\Psi ({\bf x}_a, {\bf x}_b) = 
  \int\int_{-\infty}^{\infty}\pi_a({\bf x}_a)\pi_b({\bf x}_b)d^2p.
\ee
Eq. (\ref{eigenmom2}) corresponds to a monochromatic
plane wave moving in the $-{\bf p}$ direction. 

Since $t_2 > t_1$ and also assuming that
Alice's measuring apparatus is arbitrarily large, we can ensure that
the A twins of all photons observed by Bob have already been observed by
Alice. Now after Alice's measurement B no longer necessarily 
exists in a pure state. Each particle in it may have collapsed
 to different momenta in the $xy$ plane. However the directional and
spectral filters in front of Bob's double slit as shown in Figure 1
ensure that only horizontal momentum eigenstates within a narrow bandwidth
centered on wavelength $\lambda$
are incident upon the double slit. As a result, $all$ of the ``hits" on
Bob's screen are coincidences corresponding to a narrow spectral and
angular ranges in A measurements. The wavefronts have transverse 
positional uncertainty larger than the slitwidth $d$
of the double slit. Hence, they pass through both slits to interfere and
form a Young's double slit pattern on the screen in the single counts.
Beyond the double-slit, the particle is represented by
$\Psi(x_b,y_b) = \Psi_1(x_b, y_b) + \Psi_2$,
where $\Psi_i$ is the monochromatic wave spreading from slit~$i$.
The observed intensity pattern at the screen seen in the single counts is:
\begin{subeqnarray}
\slabel{p1}
I(y) &\propto& |\Psi_1 + \Psi_2|^2, \\
\slabel{p2}
 &=& {\rm sinc}^2\left(\frac{\pi apy}{\hbar D}\right)
		 \cos^2\left(\frac{\pi dpy}{\hbar D}\right), 
\end{subeqnarray}
which is the Young's double-slit diffraction-interference pattern.
Here $a$ is the slit-width, $d$ the slit seperation, and $D$ the distance
between the slit plane and the detection screen.
We note that non-measurement by Alice also
produces the double slit pattern because of the presence of the filters. 

On the other hand, if Alice measures the transverse position and 
longitudinal momentum of the photons in ensemble
A, these photons are collapsed to eigenfunctions of the form:
\be
\label{eigenpos1}
\xi_a({\bf x}_a) = \delta (y_a - y)e^{(2\pi i/\hbar )(x_a p_x)}
\ee
which represents a monochromatic 
photon wave in A moving with momentum ${\bf p}^{\prime} = (p_x, 0)$ being
localized at $y_a = y$. If $\xi_b({\bf x}_b)$ is the eigenfunction of the 
counterpart photon in ensemble B, Eq. (\ref{wavefunc1}) is written as:-
\be
\Psi ({\bf x}_a, {\bf x}_b) = 
	\int_{-\iy}^{\iy}\xi_a({\bf x}_a)\xi_b({\bf x}_b)dydp_x,
\ee
so that a photon in B is left in a state given by the eigenfunction:
\be
\label{eigenpos2}
\xi_b({\bf x}_b) = \int^{\infty}_{-\infty}
e^{(2\pi i/\hbar )([y - y_b + y_0]p_y + [x_0 - x_b]p_x)}dp_y = 
 \hbar\delta (y - y_b + y_0)e^{(2\pi i/\hbar )(x_0 - x_b)p_x}
\ee
This represents a monochromatic photon wave in B moving with momentum
$-{\bf p}^{\prime}$ 
being transversely localized at $y_b = y + y_0$. Given that the wave 
represented by Eq. (\ref{eigenpos2}) moves perpendicular to the double slit
plane, it cannot be deselected by Bob's directional filter.
Provided $y + y_0$ falls within the area of lens 1 in Figure \ref{bm},
it passes to the spectral filter. If
its frequency falls within the allowed narrowband, it is incident upon
the double slit diaphragm. Since spectral filters restrict
only longitudinal frequency,
the localization of the eigenstate (\ref{eigenpos2}) in the transverse
direction is not affected by the filter. 
Being localized in the transverse direction,
photon B can pass through at most one slit. Each photon B
diffracts through one slit or the other in L and does not self-interfere.
Because of the low spectral and 
angular uncertainty of the incident rays, the resulting single slit
pattern is observable in the single counts. Furthermore, according to the
chronology of the Alice-Bob measurement and assuming that K
is arbitrarily large, all of Bob's hits are coincidences corresponding to
a restricted transverse localization of photons in K. This ensures high
visibility in the pattern observed by Bob.

The wavefunction behind the double slit
subsequent to Alice's position measurement is:-
$\Psi(x_b,y_b) = \Psi_1(x_b, y_b)$ or $\Psi_2(x_b, y_b)$,
where $\Psi_i$ is now the wavefunction of a particle localized at 
slit~$i$ to spread out therefrom. The intensity pattern on the screen will
be determined by the classical addition of probabilities:
\begin{subeqnarray}
\slabel{x1}
I(y) &\propto& |\Psi_1|^2 + |\Psi_2|^2 \\
\slabel{x2}
     & \approx &  {\rm sinc}^2\left(\frac{\pi apy}{\hbar D}\right).
\end{subeqnarray}
This single slit diffraction pattern 
in contrast to the diffraction-interference pattern of
Eq. (\ref{p2}).  We note that the inequality of the right-hand sides of
Eqs. (\ref{p1}) and (\ref{x1}) lies at the heart of the difference in
Bob's observed patterns.

The ensembles A and B are assumed to be sufficiently
large that a finite number of photons will be found in L after Bob's
measurement.
Depending on whether Alice chooses to measure $(p_x, y)_a$ or $(p_x, p_y)_a$,
she induces a single-slit diffraction pattern or a
Young's double-slit pattern on Bob's screen, respectively.
Conversely, based on his observation
of the interference pattern, Bob instantaneously can know what quantity 
Alice has just measured. 
Thus she is able to transmit a
classical binary signal which Bob unambiguously interprets. The effective 
speed of the signal is ${\bf v}_{\rm eff} = {\bf x}_0/\delta t$, which can be 
made arbitrarily large by increasing ${\bf x}_0$ and
decreasing $\delta t$. Nonlocality allows ${\bf v}_{\rm eff}$ to exceed $c$,
since the wavefunction is postulated to collapse instantaneously at all
points in space. A possible practical realization of this idealized
superluminal quantum telegraph is discussed in Section \ref{sqt}. 

\section{On circumventing no-signalling proofs}

The no-signalling proofs 
mentioned in the Introduction would prevent Bob from extracting a 
classical signal if he makes a momentum or position measurement {\it before}
processing the B ensemble through the directional and spectral filters,
and the double slit. The processing is such that
a transverse position state is transformed to 
a particle spreading out from one slit
or the other and does not self-interfere, while the momentum state is 
transformed to a monochromatic wave spreading from both slits that eventually
self-interferes. As a result, the subsequent evolution of
the position and momentum states beyond the double slit cannot be represented
by the same evolutionary operator. But
for this, the indistinguishability of the two mixed states resulting from
Alice's position or momentum measurment would be
merely carried over to a new eigenbasis representation.

Let us consider a simplified situation in which position
and momentum are two-valued in the entangled state. The particles A and B
can exist in two possible transverse momentum componant eigenstates 
$|p_j\rangle$ ($j$ = {\rm q, r}). Alternatively the two position 
eigenstates are $|y_i\rangle$ ($i = 1, 2)$, corresponding to a particle
being found at slit $i$. In a momentum eigenstate the particle passes through 
both slits.  The entangled state in which the particles are assumed to 
exist before Alice's measurement is:
\begin{subeqnarray}
\Psi_{1,2} 
   &=& \frac{1}{\sqrt{2}}(|p_{\rm q}\rangle|p_{\rm r}\rangle - 
       |p_{\rm r}\rangle|p_{\rm q}\rangle )\\
   &=& \frac{1}{\sqrt{2}}(|y_1\rangle|y_2\rangle - |y_2\rangle|y_1\rangle ),
\end{subeqnarray}
wherein the first eigenket in each product pair refers to an A photon, the
second to a B one.
The transformation from momentum to position basis is assumed to be given by:
\begin{subeqnarray}
\label{transform}
|y_1\rangle &=& \frac{1}{\sqrt{2}}(|p_{\rm q}\rangle + |p_{\rm r}\rangle ) \\
|y_2\rangle &=& \frac{1}{\sqrt{2}}(|p_{\rm q}\rangle - |p_{\rm r}\rangle ). 
\end{subeqnarray}
The no-signaling argument can be expressed by the fact that the reduced density 
operator ${\cal D}_p$ for Bob's particles subsequent to Alice making 
a momentum measurement is indistinguishable from
the reduced density operator ${\cal D}_y$ for Bob's particles subsequent
to her making a position measurement, as seen from:
\begin{subeqnarray}
{\cal D}_p &\equiv& \frac{1}{2}(|p_{\rm q}\rangle\langle p_{\rm q}| + 
      |p_{\rm r}\rangle\langle p_{\rm r}|)\\
    &=& \frac{1}{2}(|y_1\rangle\langle y_1| + |y_2\rangle\langle y_2|)\\
    &\equiv& {\cal D}_y
\end{subeqnarray}
This shows that Bob cannot know whether Alice made a position or 
momentum measurement by observing the B photons directly (i.e, before they 
pass filters and double slit).

To simplifiy the post-double-slit scenario, we once again associate a 
two-valuedness for the wavefunction spreading from the slits. We designate
by $|p_{i{\rm q}}\rangle$ and $|p_{i{\rm r}}\rangle$ the two possible momentum
eigenstates spreading from slit $i$. The double slit modifies a momentum 
eigenstate as:
\be
\label{ptransform}
|p_j\rangle \longrightarrow |p_j^{\prime}\rangle \equiv
 \frac{1}{\sqrt{2}}(|p_{1j}\rangle + |p_{2j}\rangle);
~~~j = {\rm q,~r}.
\ee
This depicts that a momentum eigenstate impinging on the double slit diaphragm
becomes two monochromatic wavefronts spreading from the two slits. A position
eigenstate is modified to:
\be
\label{xtransform}
|y_i\rangle \longrightarrow
|y_i^{\prime}\rangle \equiv \frac{1}{\sqrt{2}}(|p_{i{\rm q}}\rangle + 
 |p_{i{\rm r}}\rangle);
~~~i = 1,~2.
\ee
This depicts that a position eigenstate localized at slit $i$ spreads out
from the same slit as a nonmonochromatic wave. Note that the 
$|p^{\prime}\rangle$'s and 
$|y^{\prime}\rangle$'s are not mutually orthogonal, and in specific not
related according to Eq. (\ref{transform}). To understand this, we must
picture what happens to a wavefront impinging on a double slit. It is
collapsed to a wavepacket with the momentum distribution such that it peaks
at the two slits. As a result, the transformations (\ref{ptransform})
and (\ref{xtransform}) need not be unitary.
This situation is forced on us by the fundamental way in which slits affect
the wavefunction.  For example, if we form 
$|y^{\prime}_1\rangle \equiv (1/\sqrt{2})(|p^{\prime}_{\rm q}\rangle +
|p^{\prime}_{\rm r}\rangle)$ in analogy with Eq. (\ref{transform}), 
then the ket would contain
spherical momentum eigenstates originating from the other slit, which is
disallowed by the complementarity principle. 

If Alice made a momentum measurement, Bob's particles behind the double slit
are described by the density operator:-
\be
\label{pdtransform}
{\cal D}^{\prime}_p 
 = \frac{1}{2}(|p_{\rm q}^{\prime}\rangle\langle p_{\rm q}^{\prime}| +
    |p_{\rm r}^{\prime}\rangle\langle p_{\rm r}^{\prime}|)
\ee
Alternatively, in case of position measurement by Alice:-
\be
\label{xdtransform}
{\cal D}^{\prime}_y 
 = \frac{1}{2}(|y_1^{\prime}\rangle\langle y_1^{\prime}| +
    |y_2^{\prime}\rangle\langle y_2^{\prime}|)
\ee
Substituting Eqs. (\ref{ptransform}) and (\ref{xtransform}) into 
Eqs. (\ref{pdtransform}) and (\ref{xdtransform}), we find:-
\be
{\cal D}^{\prime}_p - {\cal D}^{\prime}_y =
\frac{1}{2\sqrt{2}}\left[(|p_{1{\rm q}}\rangle - |p_{2{\rm r}}\rangle)
	 (\langle p_{2{\rm q}}| - \langle p_{1{\rm r}}) \right].
\ee
The fact that ${\cal D}^{\prime}_p \ne {\cal D}^{\prime}_y$ in general implies
that Bob can know from the modified mixed states whether Alice made a 
position or momentum measurement. 

\section{A possible realization of the thought experiment}
\label{sqt}

Recent technological advances permit the performance of
experiments involving
 two-photon wavefunctions entangled in continuous variables like
momentum (Refs. \cite{strekalov,zei2000} and references cited therein). 
Typically, in such experiments a UV beam from a pump entering a nonlinear
optical crystal spontaneously fissions into two momentum entangled photons
by the process known as spontaneous parametric down-conversion (SPDC).
We describe a particular experiment closest to the thought-experiment 
described above.  An `unfolded' diagram of it is given in 
(Figure \ref{downconvert}). One of the down-converted photons
(say A) is observed by Alice scanning the focal plane $y_2$
or imaging plane $y_3$ behind a Heisenberg lens using a detector. 
The other photon (say B) from the down-conversion pair enters a (say Bob's)
double slit assembly, at distance $g$ from the crystal.
Registration of a photon A by Alice's detector at the focal plane constistutes
a momentum measurement on A. Because of
the strong correlation between the two photons, this will project photon B
to a momentum in the opposite direction.   
A diffraction-interference pattern is observed at plane $y_1$
behind the double slit in the coincidences. 
Registration of a photon A by Alice's detector at the imaging plane 
constistutes a position measurement on A. This will project photon B
to a corresponding position states As this can give
path information of photon B, it cannot pass through both slits
simultaneously and hence does not self-interfere. No double slit interference
pattern is observed in the coincidences. 

In our variant of this experiment, 
the distance $g$ is large enough that the crystal subtends a very
small angle at the double slit, and wavefronts intercepted by the latter
can be approximated by a single mode. Thus, the large distance acts as a
directional filter. Alternatively, the tandem lens system described in
Section \ref{thought} could also be used.
Alice has the choice between making a momentum measurement on photons A
by intercepting them at the focal plane, or making a position measurement
by intercepting them at the imaging plane.
\begin{figure}
\centerline{\psfig{file=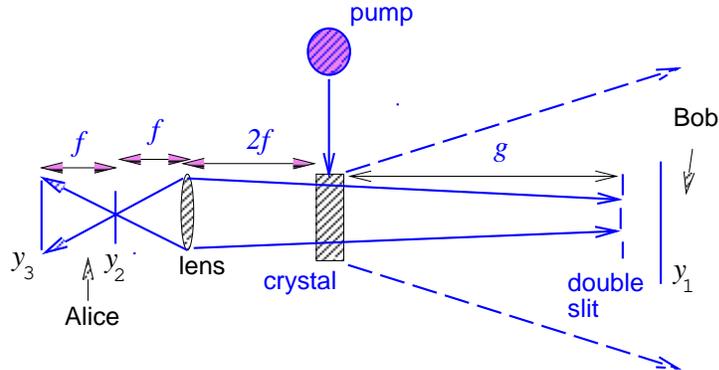}}
\caption{`Unfolded' view of a two-photon experiment:
A nonlinear crystal creates a pair of entangled
optical photons by down conversion of a UV photon from the pump.
Photon A is registered by means of a ``Heisenberg" detector system by Alice.
Photon B passes into a double slit assembly. If the detector registers a
photon at the focal plane $y_2$ of the lens, photon B is projected into a
momentum eigenstate. Bob observes a double slit pattern in the coincidences.
If the detector registers a photon at the 
imaging plane $y_3$, photon B is projected into
a position eigenstate. It can pass through at most one slit or
the other and hence no interference pattern is seen in the coincidences.}
\label{downconvert}
\end{figure}
 
If Alice registers photon A in the $y_2$ plane, photon B is left with a
well defined momentum but the position of origin at the crystal is
uncertain (Figure \ref{mom}). The corresponding photon B is instantaneously
collapsed to the modes $j$ and $k$ representing a coherent momentum state
proceeding from points $J$ and $K$ in the crystal. In practice, the situation
is more complicated since the entanglement angle of the down-converted light is 
usually finite \cite{joo96}, but we can ignore it for our purpose. 

The state $|\Psi\rangle$ of the SPDC light is assumed to be given by 
the superposition of the vacuum state and a twin state with a single photon in 
the two modes $\mu$ and $\nu$ of the signal and idler beams:
\be
\label{spdcfield}
|\Psi\rangle = 
|{\rm vac}\rangle + \epsilon(|s_\mu i_\mu\rangle + |s_\nu i_\nu\rangle ) .
\ee
where $\epsilon \ll 1$ is proportional to the classical pump field and the
crystal's nonlinearity. Here $s_X$ stands for the signal beam on ray
$X$ from the source, received by Alice; $i_X$ stands for the idler beam on
ray $X$, received by Bob.
\begin{figure}
\centerline{\psfig{figure=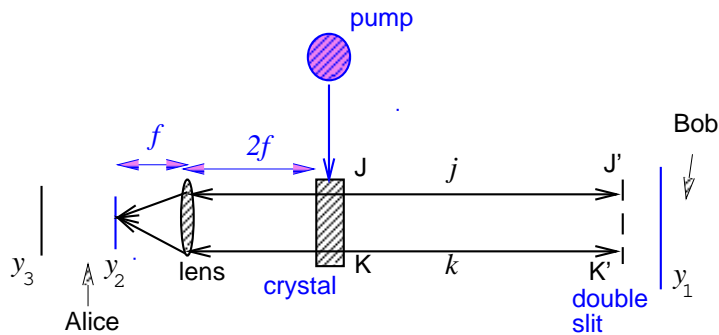}}
\caption{Simplified view of Alice's momentum measurement:
If Alice detects a photon in the $y_2$ plane she collapses
the photon B to a momentum eigenstate, and thus the direction in
which the photon B leaves the crystal is well defined. However,
the position is uncertain since the photon could have originated
at points $J$ or $K$. The momentum eigenstate passes through Bob's
double slit to self-interfere and form a double slit interference
pattern. Because of Bob's low acceptance angle, the momentum
eigenstate intercepted by him can be approximated to a unidirectional
wave. Hence the interference pattern will be visible in the single
counts.}
\label{mom}
\end{figure}

In Figure \ref{mom}, the field operator $E^{(+)}_{J^{\prime}}$ 
($E^{(+)}_{K^{\prime}}$) at the point $J^{\prime}$ ($K^{\prime}$) 
just in front of the slits plane, and $E^{(+)}_A$, the field
at Alice's detector placed in the focal plane of the lens are assumed
to be given in terms of annihilation operators:
\begin{subeqnarray}
\slabel{momfielda}
E^{(+)}_{J^{\prime}} &=& \hat{j}_i {\rm exp}(ikr_{JJ^{\prime}}) \\
\slabel{momfieldb}
E^{(+)}_{K^{\prime}} &=& \hat{k}_i {\rm exp}(ikr_{KK^{\prime}}) \\
\slabel{momfieldc}
E^{(+)}_A &=& \hat{j}_s {\rm exp}(ikr_{Jy_2}) + 
   \hat{k}_s {\rm exp}(ikr_{Ky_2}), 
\end{subeqnarray}
where $\hat{j}, \hat{k}$ are the annihilation operators in the $j$ and $k$
modes, with suffixes $s$ and $i$ used to refer to signal and idler photons.
The quantities $r_{XY}$ refer to the distance between points $X$ and $Y$
along the respective mode. The second-order cross-correlation function for 
the fields at the two points gives the two-photon wavefunction amplitude: 
\be
C_{AJ^{\prime}} \equiv \langle E^{(+)}_{J^{\prime}}E^{(+)}_A\rangle
 = \epsilon ~{\rm exp}(ikr_{J^{\prime}y_2}).
\ee
The angles $\langle\cdot\cdot\cdot\rangle$ stand for expectation value 
with the SPDC field given by Eq. (\ref{spdcfield}) with $\mu = j$ and
$\nu = k$. Similarly for the $k$ mode we find:
\be
C_{AK^{\prime}} = \epsilon~{\rm exp}(ikr_{K^{\prime}y_2}).
\ee
Since $r_{J^{\prime}y_2} \approx r_{K^{\prime}y_2}$, therefore
$C_{AJ^{\prime}} \approx C_{AK^{\prime}}$. Thus the B photon that is
incident upon the double slit is a coherent wavefront seemingly 
originating from Alice's detector on the $y_2$ plane, 
and has transverse positional uncertainty. Therefore, the wave passes 
through both slits and forms a double slit pattern. 

The twins of not all
photons registered by Alice are registered by Bob. If the photon is
registered by Bob behind the double slit, it should be a horizontal mode
that has passed within his narrow acceptance angle. 
Because of low angular uncertainty, the photons registered by Bob 
form a double slit pattern at the $y_1$-plane in the single counts.
We note that because the largeness of $g$ acts as a directional filter, the
pattern is seen even if Alice does not measure momentum.
We assume that the lens subtends approximately the same or
larger angle at the double slit than the crystal. Thus, if Alice scans
the focal plane, all registrations
by Bob will have a coincident registration by Alice (though, as mentioned,
the converse is not true), corresponding to a momentum making a small angle
with the principal axis of the lens. Furthermore, Alice will find a double
slit pattern in the coincidences even though her photons did not pass
a double slit \cite{strekalov,zei2000}.

If Alice registers a photon in the $y_3$ plane by positioning the detector
there, then the twin photon B is instantaneously
collapsed to an entangled state of the position eigenstates represented by
the modes $j$ and $m$ (Figure \ref{pos}).
In this case, the point of origin of the photons at the crystal is
definite ($J$), but momentum is uncertain. If Bob registers a coincidence, then
the mode $j$ is selected, which corresponds to a definite transverse
position on the double slit. 
Photon B can at most pass through one of the slits and 
hence forms a plain diffraction pattern. 
\begin{figure}
\centerline{\psfig{file=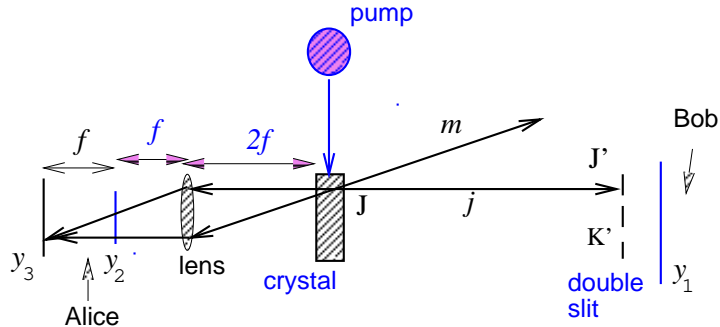}}
\caption{Simplified view of Alice's position measurement:
If Alice detects a photon in the $y_3$ plane, she collapses the
photon B to modes $j$ and $m$. The point where
the photons originated in the crystal is well defined ($J$). However, the
momentum is uncertain since the paths of the modes $j$ and $m$ have
different directions. If further the photon is detected by Bob, then mode $j$
is selected which corresponds to a well defined transverse position on
Bob's slit plane. Hence, the photon will not self-interfere to form a double
slit pattern. Because of Bob's low acceptance angle, the direction of these
photons is almost horizontal. As a result, a single slit pattern emerges
behind Bob's slit in the single counts.}
\label{pos}
\end{figure}

If Alice makes a position measurement as given in Figure \ref{pos}, then
the field $E^{(+)}_A$ at the detector placed at a point $y_3$ on the
imaging plane, and the fields at $J^{\prime}$ and $K^{\prime}$ are given by:
\begin{equation}
E^{(+)}_A = \hat{j}_s {\rm exp}(ikr_{j}) + 
   \hat{m}_s {\rm exp}(ikr_{m}) 
\end{equation}
where $r_j$ ($r_{m}$) refers to the distance between point $J$ and point
$y_3$ along mode $j$ ($m$). Using this and the detector fields given by
Eqs. (\ref{momfielda}) and (\ref{momfieldb}) we find the correlation functions:
\begin{eqnarray}
\langle E^{(+)}_{J^{\prime}}E^{(+)}_A\rangle &=& \epsilon~{\rm exp}
\left(ikr_{J^{\prime}y_2}\right), \nonumber \\
\langle E^{(+)}_{K^{\prime}}E^{(+)}_A\rangle &= &0,
\end{eqnarray}
where the averaging is done in the state given by Eq. (\ref{spdcfield})
with $\mu = j$ and $\nu = m$.
 Thus the position measurement by Alice selects a single 
longitudinal mode and localizes the photon B transversely. Since this
represents path information, photon B does not self-interfere but
instead forms a single slit diffraction pattern. Because of Bob's low
acceptance angle, only almost horizontal modes $j$ are incident upon the
double slit. Thus, Bob's time-intergrated registrations will produce the
single slit diffraction pattern in the single counts.  
Because of the angular containment of the source within the lens 
as seen by Bob, all registrations by Bob will have a 
coincident registration by Alice (as before, the converse not being true).
Thus the time-integrated pattern observed by Bob will have a high visibility.

By choosing to measure position or momentum, Alice can nonlocally control the 
single particle interference pattern seen by Bob at the $y_1$-plane
 thereby transmitting a classical binary signal. In practice, she
can be located midway between the focal and imaging plane of the lens, with
a mechanism to send light signals to activate either the detector at the
focal plane or that at the imaging plane. By making
the number flux of correlated pairs emanating from the crystal very
high the time required by Bob to integrate the received pattern
can be made sufficiently low. From Figure \ref{downconvert} 
the effective speed with which Alice transmits the signal is:
\begin{equation}
v_{\rm eff} \approx \frac{3.5f + g}{f/2c} = \left(7 + \frac{2g}{f}\right)c
\end{equation}
where the origin of the signal is taken to be Alice's position.

\section{Conclusion}

We find that nonlocality is intrinsically noncausal.
This leads us to doubt that quantum nonlocality as we understand it now can
peacefully coexist with Special Relativity. A possible test of the noncausal
effect is suggested.

\acknowledgements
I am grateful to Dr. R. Tumulka, Dr. R. Plaga, Dr. M. Steiner and 
Dr. J. Finkelstein for criticism of an earlier version of this 
manuscript. I thank Dr. Caroline Thompson for discussions.

\end{document}